\documentclass[12pt,preprint]{aastex}

\begin{document}
\title{Color Bimodality in M87 Globular Clusters}
\author{Christopher Z. Waters and Stephen E. Zepf}
\affil{Department of Physics and Astronomy, Michigan State University, 3246 
  Biomedical Physical Sciences, East Lansing, MI 48824}
\email{watersc1@pa.msu.edu,zepf@pa.msu.edu}
\author{Tod R. Lauer}
\affil{National Optical Astronomy Observatory, P. O. Box 26732, 
  Tucson, AZ 85726}
\email{lauer@noao.edu}
\author{Edward A. Baltz}
\affil{KIPAC, Stanford University, P. O. Box 20450, MS 29, Stanford, CA 94309}
\email{eabaltz@slac.stanford.edu}

\begin{abstract}
We present an analysis of a 50 orbit HST ACS observation of the M87
globular cluster system.  We use the extraordinary depth of this
dataset to test whether the colors and magnitudes show evidence for a
mass-metallicity relation in globular cluster populations.  We find
only a weak or absent relation between the colors and magnitudes of
the metal poor subpopulation of globular clusters. The weakness or
absence of a color-magnitude relation is established over a wide range
in luminosity from $M_V=-11$ to $M_V=-6$, encompassing most of the M87
globular clusters. The constancy of the colors of the metal-poor
subpopulation seen in our 50 orbit observation is in contrast to
suggestions from single orbit ACS data that the metal-poor globular
clusters in M87 and several other galaxies show a ``blue tilt.''  The
formal best fit for the mass-metallicity relation for the metal-poor
subpopulation in our much deeper data is $Z\propto M^{0.08\pm0.05}$.
Our analysis of these data also shows a possible small ``red tilt'' in
the metal-rich globular cluster subpopulation. While either of these
small tilts may be real, they may also illustrate the limit to which
mass-metallicity relations can be determined, even in such
extraordinarily deep data.  We specifically test for a wide range of
systematic effects and find that while small tilts cannot be confirmed
or rejected, the data place a strong upper limit to any tilt of
$\left|0.20\right|\pm0.05$.  This upper limit is much smaller than some
earlier claims from single orbit data, and strongly limits
self-enrichment within globular clusters. This mass-metallicity
relation for globular clusters is also shallower than the relation for
galaxies, suggesting that the formation mechanisms for these two types
of objects are different.
\end{abstract}


\section{Introduction}

The color distribution of globular clusters in most galaxies has long
been known to be bimodal \citep[e.g.][]{Kundu01,Larsen01}.  The
standard understanding of this feature is that the red and blue
subpopulations for a given galaxy represent a difference in
metallicity, with the blue clusters being on average more metal poor
than the red clusters.  The correspondence of the blue clusters with
lower metallicity and red clusters with higher metallicity follows
directly from studies of the Milky Way globular cluster system
\citep[e.g.][]{AZ98,Harris91}, and has been confirmed in a number of
extragalactic globular cluster systems
\citep[e.g.][]{CBC03,SBB07,KZ07}.  

Recently, several studies of the globular cluster systems of early
type galaxies have suggested a relation between the color and
luminosity of the blue metal-poor clusters, with brighter clusters
appearing more red \citep{Harris06,Strader,Mieske}.  This ``blue
tilt'' has been taken as evidence of a relation between the mass and
metallicity of these clusters. If such a trend were real, it would
suggest an important role for self-enrichment in luminous globular
clusters \citep{StraderSmith08}. It might also suggest that globular
clusters are surrounded by more massive halos when they form, in order
to allow them to retain metals from multiple generations of stars.
Alternatively, the absence of a mass-metallicity relation would
strongly limit the degree to which self-enrichment is a factor in the
observed cluster metallicities. Such an absence also emphasizes the
differences between the formation histories of globular clusters and
galaxies, which are known to have a strong mass-metallicity relation.

Firmly establishing any such relation requires determining the color
and luminosities of the globular clusters over a large range in
brightness to detect any effect over a significant luminosity and mass
scale.  Unfortunately, many of the results that show evidence of this
blue tilt do so using fairly shallow single orbit ACS data, with
comparably low signal to noise for the remaining more distant
galaxies.  In this paper, we present the results of an extraordinarily
deep 50 orbit ACS study of the central region of M87.  This deep data
allows clusters to be detected over the entire range of cluster
magnitudes, providing a sample that is effectively complete.  No trend
between the mass and metallicity is observed for the blue globular
cluster population, suggesting that the previously published results
are influenced strongly by the much lower signal to noise in that
data.  Section \ref{sec: obs} discusses the data and the photometric
reduction used to ensure accurate cluster measurements.  Section
\ref{sec: kmm} reviews the KMM algorithm used to determine the peaks
of the color distributions, and the sensitivity of the results to
changes in the binning methods.  Finally, section \ref{sec: results}
presents the results for our sample of clusters, and section \ref{sec:
  discussion} discusses these results in the context of the formation
history of the globular cluster system.

\section{Observations and Reductions}
\label{sec: obs}

The data for this project come from a 50 orbit observation program
with the Advanced Camera for Surveys (ACS) aboard HST (PI: Baltz,
Proposal ID: 10543).  The images are of the central region of the
giant elliptical galaxy M87, extending out to a projected radius of 8
kpc (taking $m - M = 31.021$ \citet{Macri}).  The data were
taken over the course of three months, as part of a search for
microlensing events, which require repeat visits to find changes in
brightness.  These multiple images of the same field yield data that
can be combined into very deep exposures.  On each observing day, four
exposures in F814W were taken with dithered pointing along with a
single matching exposure in F606W.  The four F814W images provide a
fully dithered image for each day, with the F606W yielding full
coverage over the entire set of observations.  The images were
combined using Multidrizzle \citep{FH} to a resolution of
$0\farcs05$ pixel$^{-1}$, the nominal resolution of ACS.  Although the
dither pattern for this dataset is more than sufficient to allow
higher resolution final images to be constructed, this is not
necessary for the identification and photometry of the globular
clusters.  At the resolution used, the globular clusters are
significantly broader than the ACS PSF, so retaining the nominal ACS
resolution provides the highest possible signal to noise.  In all, 49
F606W and 205 F814W images were combined to yield final images with
exposure times of $t_V = 24500$ s and $t_I = 73800$ s, making these
some of the deepest images ever taken with HST.  In addition to the
exposures used, 8 F606W and 13 F814W were taken but excluded from
analysis due to a loss in the guide star tracking.

The final drizzled image contains the strongly varying galaxy light.
As the main source of noise in the final image is due to the
variations in this galaxy light from pixel to pixel, constructing an
accurate model of the galaxy is essential to estimating the detection
efficiency across the image.  We use a model of the galaxy determined
from isophote fitting, but to optimize this fit, it is advantageous to
remove sources other than the galaxy light.  In addition to the rich
globular cluster population, these sources include background galaxies
and the optically bright jet emanating from the galaxy core.  In order
to identify these sources, we first create a model of the galaxy by
median filtering the image, with a box size chosen to be larger than
the expected size of any features on the image (taken to be $100
\times 100$ pixels for this resolution).  This galaxy model does a
poor job near the galaxy center, where the filtering box is unable to
deal with the steep light profile.  In addition, bright globular
clusters tend to bias the model, leading to oversubtraction around
such objects.  Despite these issues, the model subtracted data image
is significantly smoother than the initial image.  This image is then
used to construct a mask to remove the effect of the globular clusters
and jet.  All pixels that are $3\sigma$ above the local mean
(calculated again in a $100 \times 100$ pixel box) are added to the
mask image.  This method does an excellent job of cleaning the image
of the contribution from sources other than the galaxy light.  The
STSDAS ELLIPSE and BMODEL tasks were then used to create a smooth
model of the galaxy, based on the original and mask images.  The mask
ensures that bright clusters do not bias the ELLIPSE fitting, and that
the galaxy core profile is not influenced by the jet.  ELLIPSE only
runs for annuli which are more that 50\% complete (not masked or off
the frame).  To rectify this, additional C code was written to
calculate a continuation of the ELLIPSE model into the image corners,
under the assumption that the position angle and ellipticity are fixed
at the last value found by ELLIPSE.  This final model was then
subtracted from the original drizzled image, along with a constant
background level, calculated to be the mode of the galaxy subtracted
image histogram.

Source Extractor \citep{Bertin} was used to generate a database of
objects detected on the images.  To ensure that the detection
efficiency was similar across the image, the light from the galaxy was
subtracted and used to weight the detection process.  As the noise on
the final data image is highly dependent on the galaxy flux, a
constant detection threshold would miss many objects near the center
of the galaxy.  This weighting scales the detection threshold to more
accurately reflect the local image noise.  A detection threshold of
$3\sigma$ was used for this project, with a minimum area for objects of
2 pixels.  This area criterion helps limit small noise spikes from
being detected as potential clusters.  To resolve this issue further,
we also set the requirement that an object must be detected on both
the F606W and F814W images to be considered for further analysis.

All objects that were detected in our images were measured in ten
apertures from $r = 1$ pixel to $r = 10$ pixels.  As the clusters are
resolved in our images, the differences in the sizes of clusters
of equal brightness can greatly change the light detected within a
fixed aperture.  Therefore, a size dependent aperture correction was
constructed to accurately measure the total cluster flux.  To
calibrate such a correction, we created simulated globular cluster
images by convolving two dimensional \citet{King66} models with the
instrumental PSF, calculated using the empirical ACS PSFs of
\citet{AKpsfs}.  These simulated clusters were generated with a fixed
concentration $c = 1.03$ over a grid of apparent instrumental
magnitudes and tidal radii.  At each grid point, 200 simulated
clusters were randomly added to the galaxy subtracted images, and then
processed through the same detection and measurement process as for
the data images.

Based on this large sample of simulated clusters, we adopted a radius
of 4 pixels to measure the cluster light, and then determine an
aperture correction from this radius to infinity.  This correction was
parameterized by an estimate of the size of the object, defined to be
the difference in the measured magnitudes within two radii:
$\mathcal{R} \equiv m_{\mathrm 4 pxl} - m_{\mathrm 2 pxl}$.  The logic
behind this parameter is that a very small object (with a radius
smaller than 2 pixels) will have approximately the same magnitude in
both apertures, yielding a value of $\mathcal{R} \sim 0$.  As an
object increases in size, more light is measured in the large aperture
compared to the smaller, which pushes $\mathcal{R}$ to more negative
values.

The grid of simulated clusters allows for the easy construction of an
aperture correction.  For each grid point in $(m, r_t)_{simulated}$,
the median $m_{4 pixel}$ and $\mathcal{R}$ are found among the set of
simulated clusters detected by Source Extractor.  These values are
used as samples of the aperture correction:
\begin{equation}
  \mathcal{A}(\mathcal{R},m_{4 pixel}) = m_{simulated} - m_{4
    pixel}
\end{equation}
The upper end of this correction is fixed, such that
$\mathcal{A}(0,m_{4 pixel}) = 0$.  For each real object detected in
the galaxy subtracted data image, the correct aperture correction is
found by interpolating on the grid of $(\mathcal{R},m_{4 pixel})$.
This aperture correction is mainly a function of $\mathcal{R}$, with
very little dependence on the cluster magnitude.  This is exactly what
is expected, as any magnitude dependence would suggest that the
photometry is not uniform.  The median aperture corrections in our two
filters are $-0.61$ and $-0.71$ for F606W and F814W respectively, and
vary directly with the size of the cluster.

The accuracy of these aperture corrections can be checked for the data
clusters by fitting the images with PSF convolved King models
\citep{King66}.  Each of the clusters in this sample had such King
models fit to their F814W image, which has higher signal to noise than
the F606W data.  The exact details of this procedure will be presented
in a future paper \citep{WZ08c}.  The aperture correction correlates
well with the best fitting tidal radius of the clusters
\begin{eqnarray}
  \mathcal{A}_{F606W} = -0.60 - 0.0013 \times r_t \nonumber \\
  \mathcal{A}_{F814W} = -0.70 - 0.0014 \times r_t \nonumber
\end{eqnarray}
which illustrates the importance of including these size effects. The
magnitudes of the best fitting King models provide a more detailed
measurement of a given cluster's total flux.  Over the range of
magnitudes where this fitting is able to faithfully fit the cluster
images ($I \lesssim 25$), the aperture magnitudes and the King model
magnitudes agree well, with $m_{aperture} - m_{King} = -0.02 \pm
0.12$.  This scatter is consistent with our expected photometric
uncertainties, and illustrates that this aperture correction strategy
is a robust way to measure the light from the clusters with a wide
range of sizes.

The PSFs used in the generation of the simulated clusters extend out
to $0\farcs5$ from their core \citep{AKpsfs}.  The ACS detector is
known to have large angle scattering beyond this limit, which will not
be accounted for, even with this radius dependent aperture correction.
Because of this scattering, additional $-0.088$ and $-0.087$ magnitude
corrections were applied to the F606W and F814W instrumental
magnitudes, respectively, to correct them to an infinite aperture
\citep{Sirianni}.  To compare this data to previous studies of
globular clusters, we converted the instrumental F606W and F814W
magnitudes to standard V and I.  This was done using the color
correction formulas presented by \citet{Sirianni}.  The final
magnitudes were then corrected for extinction ($-0.074$ and $-0.043$
for V and I), using the extinction maps of \citet{Schlegel}.

This ACS pointing contains the majority of the clusters imaged by the
\citet{Kundu99} and \citet{Waters06} WFPC/2 studies, allowing us to
test our photometric consistency by comparing to those previously
published.  For the \citet{Kundu99} sample, there are 968 clusters
matched between the two samples with magnitude differences
\begin{eqnarray}
  V - V_{Kundu} &=& -0.048 \pm 0.127 \nonumber \\
  I - I_{Kundu} &=& 0.030 \pm 0.115 \nonumber \\
  (V - I) - (V - I)_{Kundu} &=& -0.022 \pm 0.116 \nonumber
\end{eqnarray}
This comparison uses a reanalysis of the \citet{Kundu99} sample that
incorporates the most recent official WFPC/2 zeropoints.  For the
\citet{Waters06} sample, there are 886 clusters in common, with
differences
\begin{eqnarray}
  V - V_{Waters} &=& 0.073 \pm 0.098 \nonumber \\
  I - I_{Waters} &=& 0.082 \pm 0.084 \nonumber \\
  (V - I) - (V - I)_{Waters} &=& -0.000 \pm 0.118 \nonumber
\end{eqnarray}
The original published \citet{Waters06} measurements included a $0.1$
magnitude offset in the final V magnitudes due to an inadequate PSF
used in the aperture correction.  The differences presented here
correct this error, and shows that this sample is generally consistent
with these new measurements.  For both of these samples, all bright
clusters that fall within the ACS field of view are detected.

Although this data is extraordinarily deep, and the detection
threshold has been set at a fairly liberal level, we still expect that
we are not 100\% efficient at detecting clusters.  Conveniently, the
simulated clusters used to construct the aperture correction also
sample the completeness of objects in both filters.  The fraction of
clusters detected at each grid point in apparent instrumental
magnitude and tidal radius was used to generate the expected
completeness of the data objects.  Although the detectability of an
object is a function of the object's surface brightness, there is
little dependence on the simulated cluster size, so for all further
analysis of the completeness, we consider it only as a function of the
apparent magnitude.  As expected from our simple noise model of the
galaxy, the detection efficiency, and hence the completeness, is a
function of projected radius from the center of the galaxy.  Due to
the scaling of the detection threshold, clusters with low $R_{gal}$
must have a larger flux peak to be detected at the same threshold as a
cluster with larger $R_{gal}$.  To minimize the effect this has on our
calculated completeness values, we divide the total sample into two
radius bins, separated at the median cluster radius, $68\farcs95$.
The completeness is calculated independently for the two bins, which
allows the completeness in the outer bin to extend to slightly fainter
levels.  Due to the requirement that objects be found on both the
F606W and the F814W images, the final completeness levels are color
dependent.  For the range of colors we consider for our globular
cluster sample, the median 50\% completeness limit follows the trend
$V_{50\%} \sim 25.4 + 0.8 (V - I)$.  For the following analysis of the
cluster bimodality, we use a stricter I band 90\% limit, which follows
$I_{90\%} \sim 24.5 - 0.1 (V - I)$.

The cluster detection method used was designed to detect as many
objects as possible.  Unfortunately, this also means that some
fraction of the objects found are not true globular clusters.  The two
main sources of contaminating objects are background galaxies and
noise fluctuations from the galaxy light due to random local
overdensities in the stars that make up the galaxy.  We have chosen to
make a series of cuts in the measured parameters that serve to exclude
as many of these contaminating objects as possible, while retaining
the real globular clusters.  In addition to the requirement that
objects be detected in both filters, we also apply a color cut such
that only objects with $0.5 < V - I < 1.7$ are considered to be
globular clusters.  These limits are based on the observed colors of
globular cluster systems and bracket the obvious feature in the color
magnitude diagram of the sample of all detected objects.  Next, we
only wish to consider objects that have a final completeness value
greater than 50\%.  Due to the fact that this completeness limit is so
faint, we expect that only a few real clusters will be excluded by
this cut.  Since globular clusters are known to be round, we can place
a cut on the ellipticity of the clusters in both filters, such that
$\left.\epsilon\right|_{V,I} < \frac{1}{3}$.  This cut serves as an
excellent way to exclude the many background spiral galaxies that are
seen edge on.   Finally, we exclude objects where the average surface
brightness is brighter than the peak surface brightness.  Visual
inspection of the objects excluded by this cut show them to be
entirely high spikes in the galaxy noise.

As a check of how well these cuts function to clean our catalog of
objects that are not globular clusters, we can apply these same cuts
to catalogs of objects that contain no globular clusters, and only
contaminating objects.  If the variations in the galaxy noise were
purely Gaussian, then the number of expected noise maxima on the data
image should match the number of minima.  By multiplying the galaxy
subtracted image by $-1$, and searching for objects on that image, we
should be able to estimate the noise contribution.  Unfortunately, the
asymmetry of the image histograms for these images suggest that the
fluctuations may not be Gaussian distributed, but are rather biased
towards the positive fluctuations, as might be expected from the
incipient detection of the brightest individual stars in M87.

The contamination by background galaxies obviously contributes to the
number of objects found.  A quick look at the galaxy subtracted image
shows a large number of background spiral galaxies that are easy to
identify.  The Hubble Ultra Deep Field \citep{UDF} provides a
dataset containing only background galaxies, and as such, gives a
straightforward way to estimate the contribution we expect to find
from these galaxies.  The only complication is ensuring that the data
quality matches between the two datasets.  We first rebinned the UDF
images from a pixel scale of $0\farcs030$ to the pixel scale of our
data ($0\farcs05$).  Next, the UDF was overlaid on our image
footprint, and trimmed to match the field of view.  With the geometry
matched, we considered the image noise.  The UDF count rate was
scaled to match the exposure time of the M87 data.  The M87 galaxy
model was then used to generate a noise model, which assumes that each
pixel is drawn from a Gaussian distribution with mean zero and
variance equal to the galaxy model value.  This noise was then scaled
such that when added to the UDF image, the final image variance equaled
that of the galaxy subtracted image.  This method makes it certain
that objects in the UDF frame that fall ``near the center of the
galaxy'' have more noise than objects ``at the edge.''  The reduction
for objects detected in the UDF images is identical to the M87 images,
with the only exception being that the UDF filters are F606W and
F775W.  Since these filters can be converted to the same standard V
and I system, the effect of this difference is minimal.

The numbers of objects detected with each of the various cuts in the
M87 data, the inverse images, and the UDF sample are given in table
\ref{tab: cuts}.  It is clear that the various restrictions on the
data remove the majority of the contaminating objects, and yield a
final number of clusters consistent with what is expected based on
earlier surveys.  The one object that is manually excluded from the
sample is the HST-1 knot of the jet, which due to its brightness and
compact size, successfully eludes these cuts.

\placetable{tab: cuts}

\section{Cluster Bimodality}
\label{sec: kmm}

\begin{figure}
  \plotone{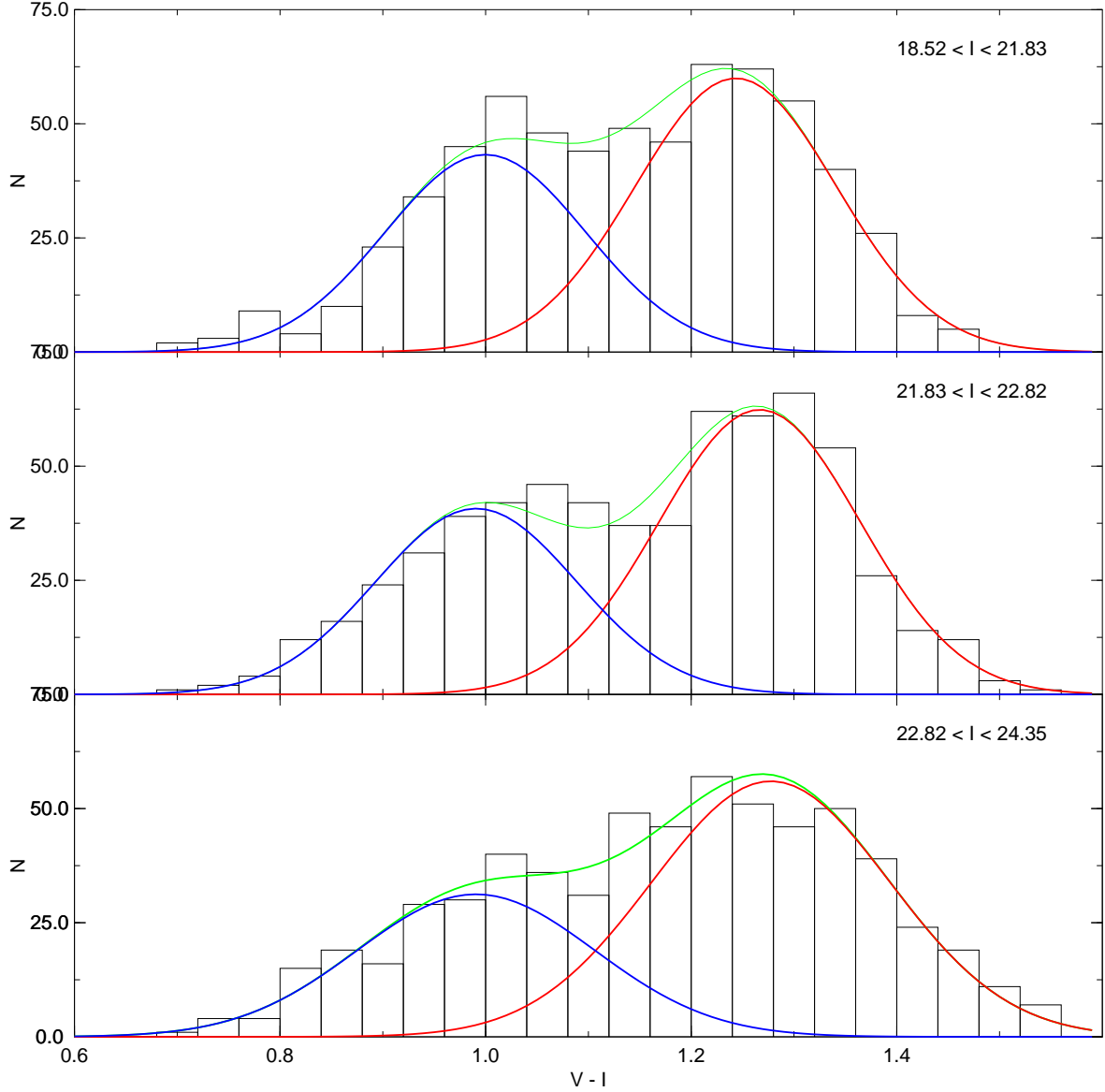}
  \caption{Histograms of the cluster colors in three bins of
    magnitude, each containing 632 clusters, along with the best
    fitting KMM component models for the red and blue populations as
    well as the sum of these two components (shown in green).  The top
    panel shows the brightest clusters, with $I < 21.94$.  The middle
    panel shows fainter clusters, with $21.94 < I < 23.06$.  The
    bottom panel shows the faintest clusters.  The constancy of the
    peaks in the red and blue populations indicate the absence of any
    strong tilts in the M87 globular cluster system.}
  \label{fig: histograms}
\end{figure}

The color distribution of our extremely deep sample of globular
clusters is shown in figure \ref{fig: histograms}.  This distribution
is clearly bimodal, as has been observed before for M87 and appears to
be a common feature of globular cluster populations
\citep[e.g.][]{Kundu01,Larsen01}.  The degree of bimodality is
generally quantified using the KMM test, as presented by
\citet{ABZ94}.  Briefly, this test determines the best fitting mixture
of Gaussian distributions to a set of data points.  For each Gaussian
component included in the mixture model, the best fitting mean and
variance are found, along with the number of points that belong to
each component.  This algorithm works quickly, using a maximum
likelihood method to estimate the best fitting parameters based on
only simple initial values.  In addition, as the data are not binned
in any way before the analysis, there is no uncertainty due to the
choice of bin size.  Closely separated groups, such as from a sample
that does not appear bimodal visually, can often be disentangled given
a sufficient number of data points (see \citealp{ABZ94} for details).

The KMM test can fit any number of Gaussian components, under the
assumption that all variances are the same (homoscedasticity) or that
the variances are possibly different (heteroscedasticity).  For our
analysis of the M87 globular cluster system, we consider a maximum of
two components, based on the obvious visual bimodality.  However, we
compare against a single component model to check the possibility that
the data is better fit by this more simple model.  This check is done
using the likelihood ratio test which is defined as
\begin{equation}
  \lambda = \frac{L(\theta|x)_{simple}}{L(\theta|x)_{complex}}
\end{equation}
This provides a quantity between zero and one that tells how likely
the simpler model fits the data better.  By comparing $-2 \ln \lambda$
to a $\chi^2$ distribution with a number of degrees of freedom equal
to twice the difference in the number of free parameters between the
two models, we can determine the point at which the more complicated
model is a sufficient improvement over the simple model to justify the
additional free parameters.  Therefore, using this test, we can check
that our data is truly fit better by the two component model by
checking that the p-value for the unimodal model is small.  

One limitation of the KMM method is that outliers in the
distribution can significantly bias the results, especially for the
heteroscedastic fits \citep{ABZ94}.  These models are biased the most
as they can adjust the individual dispersions to larger values that
better accommodate the outliers.  This in turn will tend to increase
the separation of the means of the two modes.  As the dispersions of
the homoscedastic models are coupled, the weight of an individual
outlier is significantly decreased, so many more outliers are needed
to create such an effect.  

Given that heteroscedastic fits are more influenced by outliers, it is
preferred to fit any population that does not obviously have different
dispersions with homoscedastic models.  The KMM test does not clearly
state whether the input data is better fit by a homo- or
heteroscedastic model.  Bootstrapping the sample can provide an
answer, as the value of $-2 \ln \lambda$ (as given above, using the
likelihood of the homoscedastic and heteroscedastic models for the
simple and complex cases, respectively) calculated for each
bootstrapped sample approximates the underlying $\chi^2$ distribution
\citep{Lo08}.  Therefore, the fraction of bootstrap samples with $-2
\ln \lambda$ more extreme than that calculated from the observed
data is an estimate of the p-value that the data is truly better fit
by a heteroscedastic model.

To determine whether our data are better described by a homo- or
heteroscedastic model, we fit the cluster sample using both models and
bootstrapped as described above to estimate $p_{heteroscedastic}$.
The sample used in this fit and subsequent analysis is restricted to
those clusters with color between $0.7 < V - I < 1.5$ to limit the
effect of the largest outliers.  This sample of clusters was then fit
using both models and bootstrapped as described above to determine the
probability that the data is more accurately described by a
heteroscedastic model.  The resulting dispersion for the homoscedastic
model is $\sigma_{homoscedastic} = 0.105$ and for the heteroscedastic
model $\sigma_{blue} = 0.126$ and $\sigma_{red} = 0.090$.  The
similarity of these dispersions is supported by the our statistical
test, which gives $p_{heteroscedastic} = 0.03$, excluding the
heteroscedastic fit for the entire sample at the 95\% confidence
level.  This does not mean that the underlying color dispersions of
the red and blue subpopulations are necessarily identical, only that
there is no statistical support for fitting the distributions with
different dispersions.  Therefore, the remainder of this paper focuses
on the results for the homoscedastic models, which we again note are
much more robust against the influence of outliers.

\begin{figure}
  \plotone{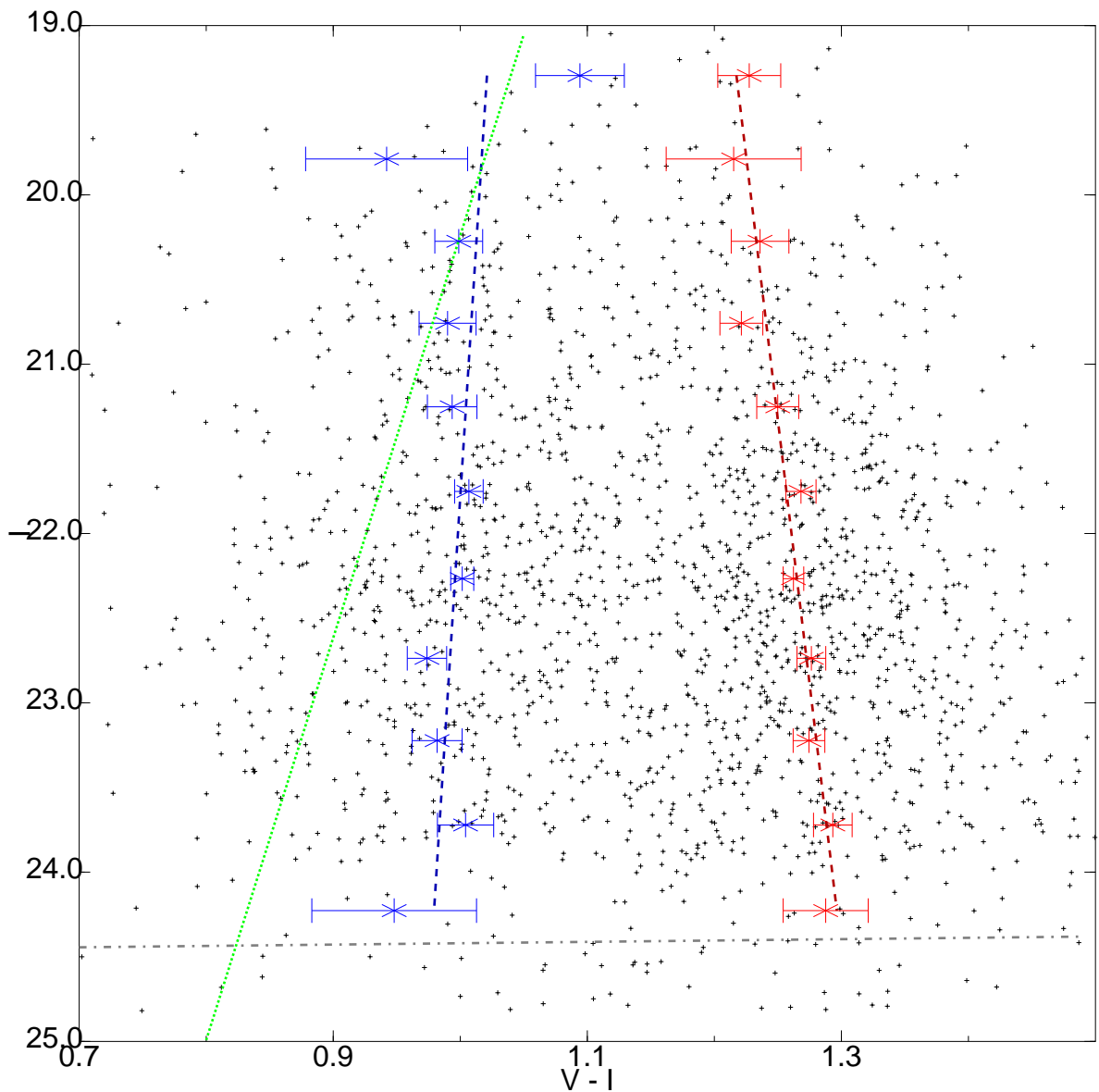}
  \caption{Color trends with I magnitude for the globular clusters in
  these data, based on 0.5 magnitude width bins with homoscedastic fits.  The errors are calculated via bootstrapping.  The best 
  fitting trends show effectively no blue tilt ($Z \propto M^{0.08 \pm 0.05}$),
  and is significantly different than the previously published trend
  $Z \propto M^{0.55}$ shown as a dotted line.  For reference, the
  median 90\% completeness line is shown as a dot-dashed line.
  \label{fig: fixed bin 0.5m}}
\end{figure}

As we wish to investigate any trends in bimodality with cluster mass,
we need to divide the clusters into luminosity bins.  Following
previous studies \citep{Strader,Harris06}, we use the I magnitude as a
tracer of cluster mass.  Two types of magnitude bins were used for
this study.  The first method uses fixed width bins from $I = 19.5$ to
$I = 24.5$, using bin widths of 0.5 and 1.0 magnitudes.  The mean
value of I is calculated for each bin, which due to the shape of the
globular cluster luminosity function, does not fall exactly at the
center of the bin range.  Unfortunately, the KMM test as originally
stated does not directly yield uncertainties for the parameters
calculated.  Since the algorithm converges quickly, it is reasonable
to estimate these uncertainties by bootstrapping the sample
\citep{Basford97}.  Therefore, the best fitting means, variances, and
population ratios were calculated with their associated uncertainties
from 100 bootstrapped samples in each magnitude bin.  These values are
listed in \ref{tab: fixed bin 0.5m} and \ref{tab: fixed bin 1.0m},
with the results for the 0.5 magnitude bins plotted in figure
\ref{fig: fixed bin 0.5m}. 

The second binning method uses running samples of 100 clusters.  The
clusters are sorted in I magnitude, and the KMM test run on the first
100 points.  For the next bin, the brightest cluster in the sample is
removed and the next cluster from the sorted list of magnitudes is
included, and the process repeated.  As these bins contain on average
fewer clusters than any of the fixed width bins, the number of
outlying points in any of the running bins should also be small.  The
effect of these outliers can be somewhat mitigated by bootstrapping,
as not all samples will retain the discrepant points.  However, taking
the average values from these bootstrap results allows the outlying
points to continue to influence the results.  One solution to remove
this influence is to use the bootstrap ``bumping'' procedure of
\citet{TK99}.  For each set of bootstrapped results, the fit with the
largest likelihood is retained and the rest discarded.  This makes the
fitting resistant to outliers, as the small probability of finding
such a point given the model ensures that any sample that contains the
outlying point will naturally have a much lower likelihood than a
sample that has excluded that point.  Each bin was bootstrapped in
this way with 200 samples, which should allow the fitting to be
resistant to the influence of up to 5 outliers \citep{TK99}.

The best fitting models for the fixed width bins are shown in tables
\ref{tab: fixed bin 0.5m} and \ref{tab: fixed bin 1.0m}.  We can
see that based on these models, the unimodal description
of the cluster colors is strongly ruled out over most of the range in
magnitude considered.  The only deviations from this trend occur at
the very brightest and faintest bins.  These deviations are not
surprising, as these bins have a smaller number of
clusters. Simulations by \citet{ABZ94} showed for cases like these,
with a small number of objects ($N \sim 50)$ and a modest ratio of
component mean separation to component variance ($\Delta \mu =
\frac{\mu_2 - \mu_1}{\sigma} \sim 2$), there is a high probability of
a truly bimodal distribution being unrecognized by the KKM algorithm

Based on these fits, we can see that as we look at bins containing
fainter clusters, the variances in the best fitting models increase.
This is true for both subpopulations, and shows the effect the
decrease in signal to noise has on the scatter in the measured colors.
The population fractions for the red and blue fits show that as the
bins move to fainter magnitudes, the blue population fraction falls.
This is a result of the differences in the luminosity functions of the
red and blue clusters and will be discussed in a future paper on the
globular cluster luminosity function of this very deep data (Waters et
al. 2008).

\section{Color-Metallicity and Mass-Metallicity Relations}
\label{sec: results}

We use the means determined by KMM for the red and blue clusters as a
function of I magnitude to investigate the existence of any
color-magnitude trends in both populations.  For each binning method,
the best fitting trend is calculated for both the red and blue
subpopulations.  For the fixed width bins that have errors from
bootstrapping, these errors are used to weight the fits.  This helps
limit the influence from the brightest and faintest bins, which tend
to have large errors due to the low number of clusters.  The running
bins constrain the fit with equal weight, as the large number of these
bins reduces the relative influence of any individual bin.

These fits yield the best trend $(V - I) = a + b \cdot I$, which can
be converted to a mass metallicity relation of the form
$\left[\mathrm{Fe} / \mathrm{H}\right] = k + \alpha \log_{10}
\frac{M}{M_\odot}$.  We follow \citet{Harris96} in creating a relation
\begin{equation}
  \left[\mathrm{Fe} / \mathrm{H}\right] = 5.2267 (V - I) - 6.2613
\end{equation}
based on observations of Milky Way globular clusters, and use a
constant mass to light ratio of $M / L = 3$ for all clusters.  Table
\ref{tab: best fit trends} shows the best fitting values for these
trends for each of the binning methods.  It is clear that for the blue
subpopulation, all three binning methods yield similar best fits.  To
determine our formal best fit models, we average the three binning
methods, and take the scatter in these values as our expected
uncertainty in the fit due to the binning.  Over the full range of
clusters, we find an average best fit for the blue clusters of $Z
\propto M^{0.08 \pm 0.05}$.  To offer a more direct comparison with
the tilts measured in previous studies, we have also fit only those
clusters brighter than the turnover ($I \sim 22.5$).  These fits are
given in table \ref{tab: best fit trends to}. We find a similarly
small $Z \propto M^{0.01 \pm 0.07}$ relation with these limits, and as
they are only marginally different than the fits over the full
magnitude range, we do not plot these trends separately.

It is essential to characterize how well the formal uncertainties in
the fit reflect the true uncertainty in the slope of the
mass-metallicity relation.  In order to test the accuracy of the
calculated fits, we constructed simulated color-magnitude diagrams
with a known tilt in the blue clusters.  Each cluster in our data
sample was checked against the 0.5 magnitude homoscedastic fixed
width bins to determine if it was a likely member of the blue
subpopulation.  All of these blue clusters then had their color
shifted, such that
\begin{equation}
  (V - I)_{new} = (V - I)_{old} + (a_{trend} - a_{blue}) + 
                   I (b_{trend} - b_{blue})
\end{equation}
where $a_{blue}$ and $b_{blue}$ are the best fitting trend in the blue
clusters, and $a_{trend}$ and $b_{trend}$ define the trend we are
adding.  This method therefore preserves the cluster I magnitude, as
well as the distribution of points around the best fit trend.  These
simulated color magnitude diagrams are then run through the KMM test,
and the best calculated trend found.  By comparing the difference in
the input and output slope, we can estimate how large we can expect
the errors to be in our mass-metallicity fits.

\begin{figure}
  \plottwo{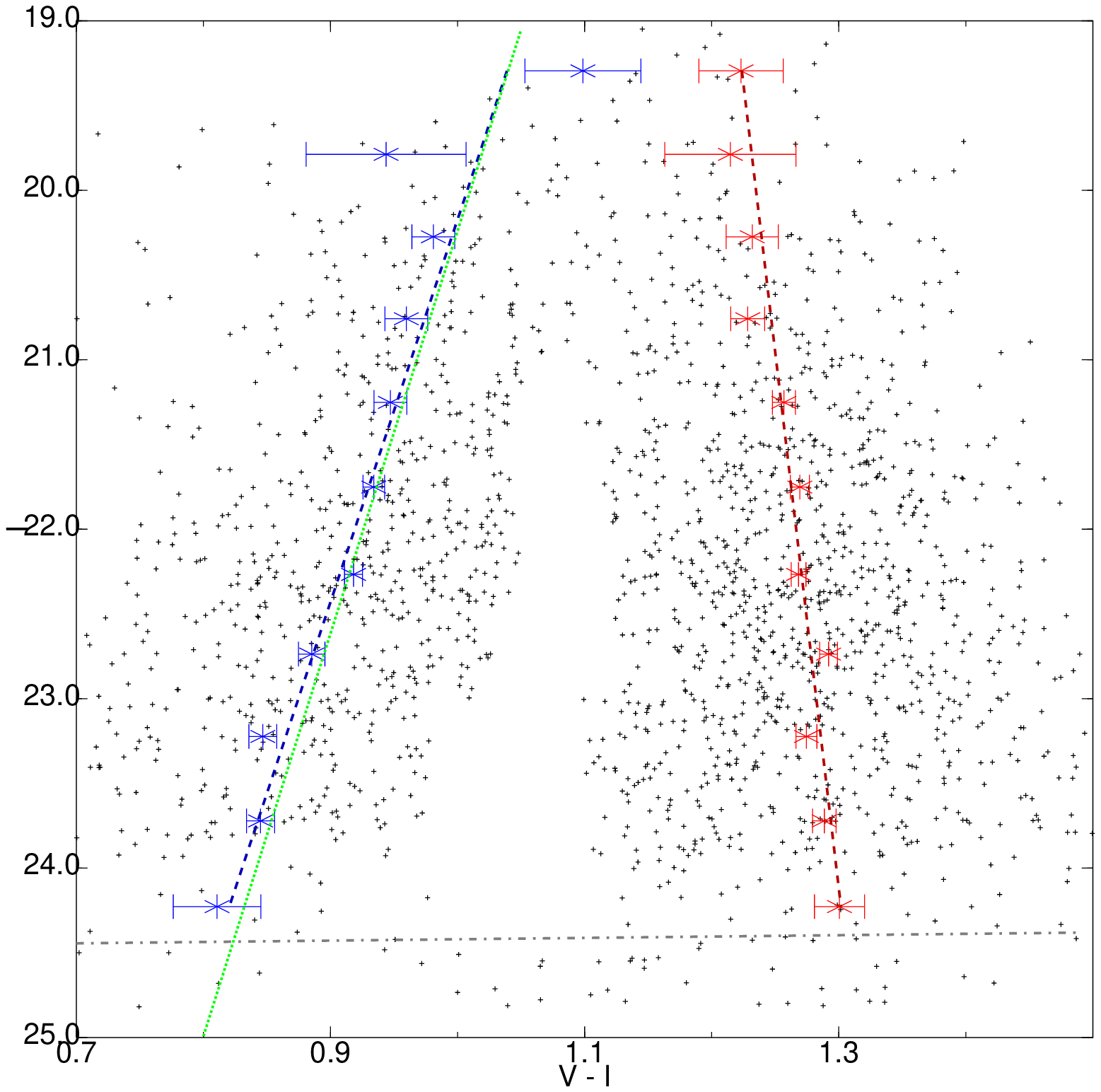}{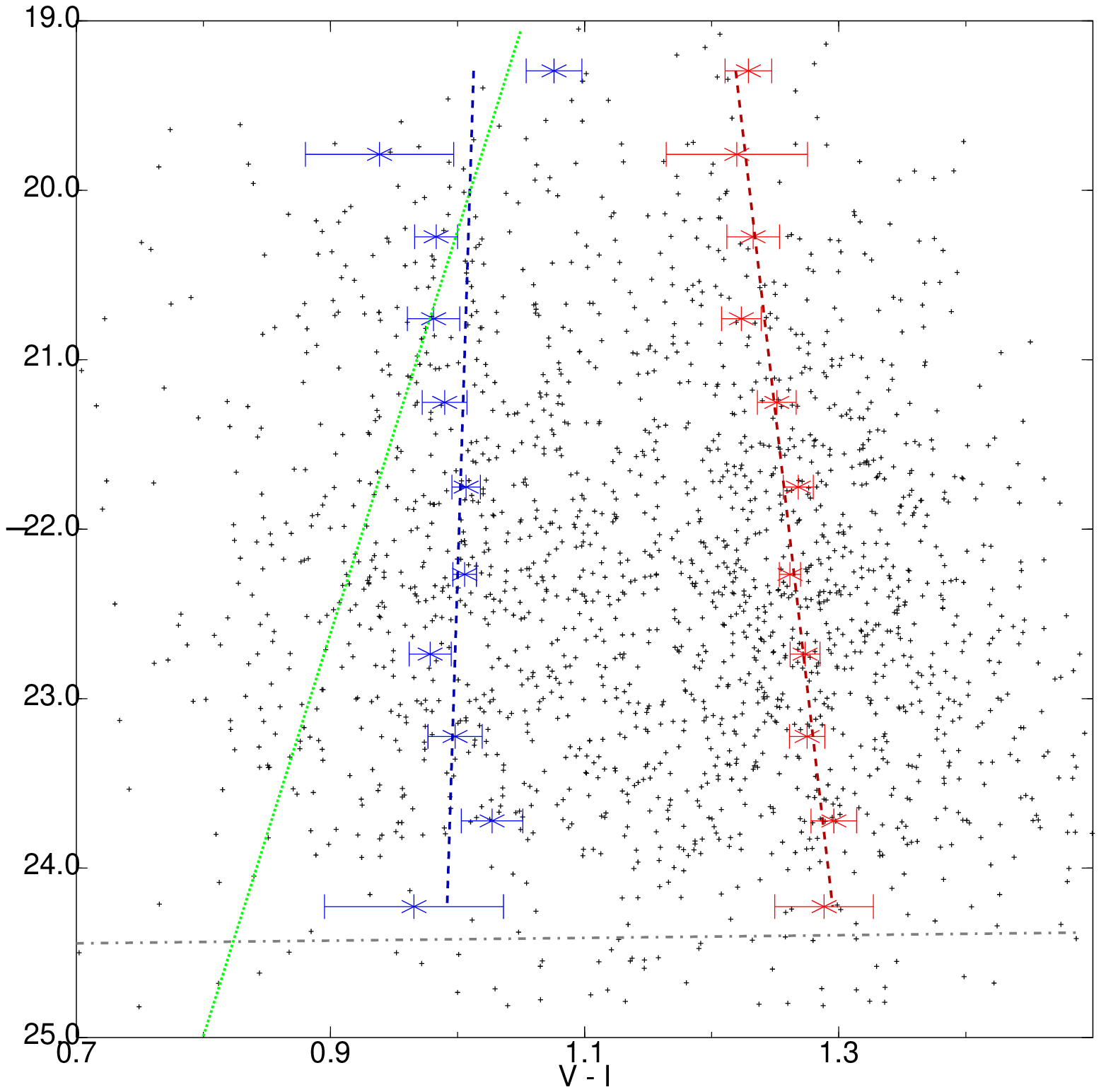}
  \caption{Color magnitude diagram of a set of simulated clusters with
    a mass-metallicity trend defined to be $Z \propto M^{0.55}$
    (left).  As shown by the best fitting line ($Z \propto M^{0.58 \pm
      0.05}$), if the clusters truly had such a trend, it would be
    easily identifiable in our data.  The right panel shows the
    results of fitting simulated clusters with no mass-metallicity
    trend at all.  The best fitting line for this data is also shown
    ($Z \propto M^{0.06 \pm 0.06}$), which falls within the
    uncertainty of our fits.
    \label{fig: simulated}}
\end{figure}

Two simulated trends were used for this test, representing both
extremes in the blue tilt.  First, a strong tilt of $Z\propto
M^{0.55}$ like that claimed in some analyses of shallower data was
added, the results of which are shown in figure \ref{fig: simulated}.
The best fitting trend for this simulation is $Z \propto M^{0.58 \pm
  0.05}$ ($Z \propto M^{0.44 \pm 0.05}$ including only those clusters
brighter than the turnover), suggesting that the small formal errors
calculated are not significant underestimates. As we recover this
large input trend with high accuracy, it is clear that were such a
trend truly present in our data, we would have no difficulty in
detecting it.  The main reason that such a large trend is so easy to
recover is that it must have a large separation between the means of
the red and blue subpopulations at faint magnitudes.  This separation
is much larger than the color dispersion of the clusters in our
sample, which creates a gap between the two groups.  Such a gap allows
the KMM test to easily group the clusters, providing an excellent fit.

The creation of a large gap is not an issue for the second simulation,
in which the slight blue trend is completely removed, with a simulated
trend of $Z \propto M^{0.00}$.  As there is only a slight shift in the
cluster colors, the degree of blending between the red and blue
clusters should be nearly identical to the real cluster data.  The
calculated trend for this simulation is $Z \propto M^{-0.06 \pm 0.06}$
($Z \propto M^{0.07 \pm 0.06}$ including only those clusters brighter
than the turnover).  These tests on simulated data indicate that our
fits to the color magnitude trends in the globular cluster
subpopulations accurately reflect the underlying trend in the data with
small well-understood uncertainties.  The specific fitting
uncertainties in the fit to the slope have a maximum of $0.06$, which
occurs for small real slopes.  This uncertainty decreases to even
smaller values when the input slope is large.  However, our method
does not directly address any possible systematic issues in the
measurement of the colors and magnitudes.  While our measurements are
based on a careful analysis of this extraordinarily deep data, it is
important to consider all possible systematic effects in the
measurement.

Our tests with artificial datasets indicate that we accurately recover
the simulated slopes with a maximum uncertainty of $0.06$.  These
tests provide an excellent way to assess the uncertainties in the fits
to the data.  However, this approach does not directly address
potential systematic biases in the measurements of the colors and
magnitudes.  The most likely source of any systematic error in these
is due to the aperture corrections.  The extraordinary depth of our
data allows a much more accurate determination of the sizes of the
globular clusters, which directly leads to a much more accurate
determination of the total magnitude of the globular clusters.  These
sizes are influenced by the choice of PSF used, and we simulated the
effects of different PSFs on the resulting aperture corrections.  Our
simulations show that in this well sampled data, in which the globular
clusters are spatially resolved, large differences in the magnitudes
(of up to $0.1$ magnitude) only occurred with PSFs that were
inconsistent with our \citet{AKpsfs} PSFs and were not good fits to
the single unsaturated star in the image.  Using such poor PSFs can
create tilts of up to $\pm 0.2$ in the final mass metallicity relation
for both subpopulations on top of the modest tilts seen in our
fitting.  This extreme allows us to note that any possible systematic
effect due to the aperture correction must be smaller than this level,
and that both of the slopes found in our data fall within this range
of $\alpha < \left|0.2\right| \pm 0.06$.

\section{Discussion}
\label{sec: discussion}

Our 50 orbit ACS observation of M87 shows no significant relation
between the colors of the blue metal poor clusters and their
luminosity, with a formal best fit of $Z \propto M^{0.08}$ from $19.5
< I < 24.5$, and a conservative upper limit on the slope of any
relation in the M87 globular cluster population of $\alpha <
\left|0.20\right| \pm 0.06$ including systematic effects.  A similarly
small trend is found when the fitting is restricted to only the bright
clusters above the luminosity function turnover ($19.5 < I < 22.5$).
This absence of any significant mass-metallicity relation rules out
some earlier claims of such a trend from much shallower data.  Some
earlier work investigated the mass metallicity relation of metal poor
globular cluster populations in much shallower images of nearly
ellipticals, including single orbit data for M87 and several other
bright Virgo ellipticals \citep{Strader,Mieske}, single orbit
pointings of NGC 4594 \citep{Spitler}, and several distant ellipticals
with longer exposures and thus similar signal to noise as the nearby
single orbit observations \citep{Harris06}.  These shallower studies
suggested that the metal poor globular cluster populations of some of
these galaxies, including M87, had a blue tilt and inferred a mass
metallicity relation of about $Z \propto M^{0.55}$.  These results are
clearly not confirmed in our 50 orbit data, which places an upper
limit on any mass-metallicity relation that is much smaller than this
strong trend.  There are also several ground based studies of the
color-magnitude trends in globular cluster systems that find a variety
of results, including both blue tilts \citep{FFG,Wehner} as well as
red tilts \citep{Bassino,Lee}. \citet{FFG} examined M87, and found
evidence for mass-metallicity relation of $Z \propto M^{0.44}$, which
although smaller than the claims from space-based studies, is still
inconsistent with the lack of a tilt in our much deeper data.  This
emphasizes the difficulty in accurately determining color tilts from
data with low signal to noise and poor spatial resolution.

The clear absence of a significant mass metallicity effect in our data
contrary to the very strong effect in single orbit data for galaxies
like M87 and in data with similar signal to noise at larger distance,
highlights the need for very deep observations to address this
question.  \citet{Kundu08} has investigated this question in detail.
We do not reproduce this extensive work here, but note that
\citet{Kundu08} identifies two major issues that arise in single orbit
data or data with similar signal to noise for the bulk of the
clusters.  First, such data lacks the depth necessary to accurately
follow the sizes of globular clusters with magnitude.  This lack of
size discrimination will cause size dependent photometric errors.
Secondly, the error bars for detections are not symmetric in color and
magnitude.  These two effects are shown to be able to produce apparent
color magnitude trends of the level seen in the lower signal to noise
data from an underlying distribution with no trend at all
\citep{Kundu08}.

The absence of a significant mass-metallicity relation for globular
clusters also suggests a fundamental difference between globular
clusters and galaxies, as galaxies have a well-known mass metallicity
relation.  Specifically, using SDSS data, \citet{Tremonti} found a
mass metallicity relation of $Z \propto M^{0.3}$ for a very large
sample of galaxies. This was extended to nearby dwarf irregular
galaxies by \citet{Lee06}, who found a mass-metallicity relation
consistent with that for the more massive galaxies. This dwarf galaxy
sample extends down to $M\sim 10^6 - 10^7$, the range where the most
massive globular clusters are found, and thus is suggestive of a
fundamental difference between globular clusters and galaxies in their
mass-metallicity relations.  Such a difference likely reflects
differences in the formation histories of galaxies and globular
clusters.  If globular clusters experience significant
self-enrichment, then the more massive clusters will appear more
metal-rich, since their greater mass will enable them to retain more
metals from earlier generations of stars.  Any self-enrichment of
metals that affect the broad-band colors will make the more massive
clusters redder, and produce color-luminosity and mass-metallicity
relations within a globular cluster system.  Therefore, the weakness
or absence of the observed color-luminosity and mass-metallicity
relations for globular clusters thus sets a limit on the role of
self-enrichment, and is a key target for future models of globular
cluster formation.

A natural explanation for the difference between the mass-metallicity
relations of globular clusters and galaxies is that globular clusters
form without extensive mass distributions or dark matter halos.
Without such halos, globular clusters are unable to retain the
material produced by their massive stars, preventing the formation of
subsequent generations of metal-enriched stars.  Such a picture is
consistent with the compact, dense nature of globular clusters which
implies short formation timescales, and with models in which globular
cluster formation is a rapid, dynamic process in a high pressure
starburst environment as suggested by observations of globular cluster
formation in the local universe \citep{AZ01,EE}.  In contrast then,
galaxies tend to form over time within larger dark-matter dominated
structures that help retain metals to be incorporated in subsequent
generations to produce the observed galaxy mass-metallicity relation.

\acknowledgements

CZW and SEZ acknowledge support for this work from HST grant number
HST-10543 and NSF award AST-0406891.  We also acknowledge useful
conversations with Arunav Kundu on possible sources of photometric
error.

\clearpage

\begin{deluxetable}{l c c c}
  \tabletypesize{\small}
  \tablewidth{0pc}
  \tablecaption{Number of Detected Objects
  \label{tab: cuts}}
  \tablehead{\colhead{Cut} & \colhead{$N_{clusters}$} &
    \colhead{$N_{inverse}$} & \colhead{$N_{UDF}$}}
  \startdata
  Matched               & 5392 & 136 & 242 \\
  Completeness $>$ 50\% & 3996 & 45  & 231 \\
  $0.5 < V - I < 1.7$   & 2832 & 13  & 110 \\
  Ellipticity           & 2168 & 4   & 31  \\
  Surface Brightness    & 2090 & 2   & 31  \\
  \hline
  Final                 & 2089 & 2   & 31
  \enddata
\end{deluxetable}

\begin{deluxetable}{c cc cc cc cc}
  \rotate
  \tabletypesize{\small}
  \tablewidth{0pc}
  \tablecaption{Best fit values: 0.5 magnitude bins/homoscedastic
  \label{tab: fixed bin 0.5m}}
  \tablehead{\colhead{I} & 
    \colhead{$\langle V - I \rangle_{blue}$} & \colhead{$\sigma_{blue}$} &
    \colhead{$\langle V - I \rangle_{red}$} & \colhead{$\sigma_{red}$} &
    \colhead{$\pi_{blue}$} & \colhead{$\pi_{red}$} &
    \colhead{$P_{unimodal}$} & \colhead{$N_{clusters}$}}
  \startdata
19.295 & 1.094$\pm$0.035 & 0.046$\pm$0.009 & 1.227$\pm$0.025 & 0.046$\pm$0.009 & 0.460$\pm$0.196 & 0.540$\pm$0.196 & 0.312 & 15\\
19.788 & 0.942$\pm$0.064 & 0.115$\pm$0.011 & 1.215$\pm$0.053 & 0.115$\pm$0.011 & 0.419$\pm$0.187 & 0.581$\pm$0.187 & 0.269 & 42\\
20.274 & 0.999$\pm$0.019 & 0.088$\pm$0.009 & 1.236$\pm$0.023 & 0.088$\pm$0.009 & 0.535$\pm$0.082 & 0.465$\pm$0.082 & 0.012 & 90\\
20.758 & 0.990$\pm$0.022 & 0.095$\pm$0.009 & 1.221$\pm$0.017 & 0.095$\pm$0.009 & 0.430$\pm$0.067 & 0.570$\pm$0.067 & 0.016 & 134\\
21.252 & 0.994$\pm$0.020 & 0.092$\pm$0.009 & 1.250$\pm$0.017 & 0.092$\pm$0.009 & 0.432$\pm$0.068 & 0.568$\pm$0.068 & 0.000 & 170\\
21.752 & 1.007$\pm$0.011 & 0.092$\pm$0.006 & 1.268$\pm$0.012 & 0.092$\pm$0.006 & 0.415$\pm$0.043 & 0.585$\pm$0.043 & 0.000 & 262\\
22.267 & 1.002$\pm$0.009 & 0.086$\pm$0.005 & 1.262$\pm$0.008 & 0.086$\pm$0.005 & 0.424$\pm$0.035 & 0.576$\pm$0.035 & 0.000 & 332\\
22.737 & 0.974$\pm$0.015 & 0.106$\pm$0.005 & 1.276$\pm$0.011 & 0.106$\pm$0.005 & 0.325$\pm$0.036 & 0.675$\pm$0.036 & 0.000 & 288\\
23.223 & 0.982$\pm$0.020 & 0.112$\pm$0.006 & 1.274$\pm$0.012 & 0.112$\pm$0.006 & 0.370$\pm$0.042 & 0.630$\pm$0.042 & 0.000 & 270\\
23.722 & 1.004$\pm$0.022 & 0.112$\pm$0.008 & 1.293$\pm$0.015 & 0.112$\pm$0.008 & 0.384$\pm$0.063 & 0.616$\pm$0.063 & 0.000 & 189\\
24.227 & 0.948$\pm$0.065 & 0.128$\pm$0.012 & 1.288$\pm$0.033 & 0.128$\pm$0.012 & 0.262$\pm$0.128 & 0.738$\pm$0.128 & 0.059 & 55\\
  
\enddata
\end{deluxetable}

\begin{deluxetable}{c cc cc cc cc}
  \rotate
  \tabletypesize{\small}
  \tablewidth{0pc}
  \tablecaption{Best fit values: 1.0 magnitude bins/homoscedastic
  \label{tab: fixed bin 1.0m}}
  \tablehead{\colhead{I} & 
    \colhead{$\langle V - I \rangle_{blue}$} & \colhead{$\sigma_{blue}$} &
    \colhead{$\langle V - I \rangle_{red}$} & \colhead{$\sigma_{red}$} &
    \colhead{$\pi_{blue}$} & \colhead{$\pi_{red}$} &
    \colhead{$P_{unimodal}$} & \colhead{$N_{clusters}$}}
  \startdata
19.658 & 0.888$\pm$0.073 & 0.111$\pm$0.012 & 1.175$\pm$0.027 & 0.111$\pm$0.012 & 0.198$\pm$0.122 & 0.802$\pm$0.122 & 0.074 & 57\\
20.564 & 0.995$\pm$0.012 & 0.093$\pm$0.007 & 1.227$\pm$0.015 & 0.093$\pm$0.007 & 0.468$\pm$0.049 & 0.532$\pm$0.049 & 0.001 & 224\\
21.555 & 1.001$\pm$0.010 & 0.093$\pm$0.005 & 1.262$\pm$0.009 & 0.093$\pm$0.005 & 0.427$\pm$0.029 & 0.573$\pm$0.029 & 0.000 & 432\\
22.485 & 0.991$\pm$0.009 & 0.097$\pm$0.004 & 1.269$\pm$0.007 & 0.097$\pm$0.004 & 0.372$\pm$0.024 & 0.628$\pm$0.024 & 0.000 & 620\\
23.428 & 0.990$\pm$0.014 & 0.113$\pm$0.005 & 1.281$\pm$0.010 & 0.113$\pm$0.005 & 0.375$\pm$0.036 & 0.625$\pm$0.036 & 0.000 & 459\\

  \enddata
\end{deluxetable}

\begin{deluxetable}{c cccc cccc}
  \tabletypesize{\small}
  \tablewidth{0pc}
  \tablecaption{Best fit trends
  \label{tab: best fit trends}}
  \tablehead{\colhead{Model} &
    \colhead{$a_{blue}$} & \colhead{$b_{blue}$} & 
    \colhead{$k_{blue}$} & \colhead{$\alpha_{blue}$} &
    \colhead{$a_{red}$} & \colhead{$b_{red}$} & 
    \colhead{$k_{red}$} & \colhead{$\alpha_{red}$}}
  \startdata
Fixed 0.5 mag &	1.184 & -0.008 & -1.749 & 0.110 & 0.908 & 0.016 & 1.666 & -0.209 \\
Fixed 1.0 mag &	1.029 & -0.002 & -1.197 & 0.021 & 0.824 & 0.020 & 1.972 & -0.258 \\
100 point running bins &	1.199 & -0.009 & -1.728 & 0.114 & 1.015 & 0.011 & 1.231 & -0.144 \\

  \enddata
\end{deluxetable}

\begin{deluxetable}{c cccc cccc}
  \tabletypesize{\small}
  \tablewidth{0pc}
  \tablecaption{Best fit trends above turnover
  \label{tab: best fit trends to}}
  \tablehead{\colhead{Model} &
    \colhead{$a_{blue}$} & \colhead{$b_{blue}$} & 
    \colhead{$k_{blue}$} & \colhead{$\alpha_{blue}$} &
    \colhead{$a_{red}$} & \colhead{$b_{red}$} & 
    \colhead{$k_{red}$} & \colhead{$\alpha_{red}$}}
  \startdata
Fixed 0.5 mag &	1.135 & -0.006 & -1.538 & 0.080 & 0.906 & 0.016 & 1.672 & -0.210 \\
Fixed 1.0 mag &	1.009 & -0.001 & -1.118 & 0.009 & 0.733 & 0.024 & 2.338 & -0.314 \\
100 point running bins &	0.928 & 0.004 & -0.596 & -0.054 & 0.794 & 0.021 & 2.118 & -0.278 \\
  
\enddata
\end{deluxetable}


\end{document}